\documentclass[12pt]{iopart}

\usepackage{epsf} \usepackage{epsfig}
\usepackage{iopams}

\begin{document}

\title {Overlap Interfaces in Hierarchical Spin-Glass models} \vskip 10pt
\author{Silvio Franz$^1$, T J\"org$^1$ and Giorgio Parisi$^2$} \address{$^1$
  LPTMS, UMR 8626 CNRS et Universit\'{e} Paris-Sud, B\^{a}timent 100, 91405
  Orsay CEDEX, France} \address{$^2$ Dipartimento di Fisica and INFM-CNR,
  Universit\`{a} di Roma ``La Sapienza'', Piazzale Aldo Moro 2, 00185 Roma,
  Italy}

\date{today}

\begin{abstract}
  We discuss interfaces in spin glasses. We present new theoretical results
  and a numerical method to characterize overlap interfaces and the
  stability of the spin-glass phase in extended disordered systems. We use
  this definition to characterize the low temperature phase of hierarchical
  spin-glass models. We use the Replica Symmetry Breaking theory to evaluate
  the cost for an overlap interface, which in these models is particularly
  simple. A comparison of our results from numerical simulations with the
  theoretical predictions shows good agreement.
\end{abstract}


\section{Introduction}
\label{sec:intro}

The study of interfaces is often used to investigate the stability of
ordered coexisting phases. Low temperature ordered phases are spatially
homogeneous, however, interfaces can be induced by forcing heterogeneous
boundary conditions \cite{ZJ}. For example, in the case of phases
characterized by different values of an order parameter, one can fix the
order parameter at the boundaries along one selected direction to two
different values. For stable phases the induced interface costs an amount of
free-energy proportional to a power of the interface size.  At the lower
critical dimension this power vanishes and long-range order becomes
unstable.

In spin glasses, due to the lack of a physical order parameter
allowing to distinguish the different phases, this definition of an
interface does not seem of practical use.  Different proposals have
been put forward to define interfaces in spin glasses.  A common
procedure \cite{Southern-Young,McMillan,Bray-Moore}, involves the
comparison of free-energies between systems with periodic and
antiperiodic boundary conditions, respectively.  This leads to a
definition of an interface exponent, from the relation: $|F_{\rm
  per}-F_{\rm anti-per}|\sim L^{\theta}$ which plays a fundamental
r\^ole in the droplet picture of spin glasses
\cite{Bray-Moore,Fisher-Huse}. Measures of $\theta$ have been used to
perform a high-precision estimate of the spin-glass lower critical
dimension $D_C \approx 2.5$ in Ref.~\cite{Boettcher} (defined as the
value of $D$ for which the interface exponent $\theta$ vanishes). An
alternative definition \cite{FPV} uses the free-energy cost that is
needed to impose a spatial heterogeneity in the order parameter, which
in spin glasses is the overlap, i.e. a measure of the similarity
between two equilibrated configurations of the same system \cite{MPV}.
One needs to consider two identical copies of the system constrained
in such a way to have their mutual overlaps on the boundaries fixed to
some preassigned values.  This definition is well suited to test an
important property emerging in the Replica Symmetry Breaking (RSB) of
the spin-glass phase: while the probability distribution $p(q)$ for
the overlap is broad\footnote{This is in contrast to the droplet
  picture of the spin-glass phase where the distribution $p(q)$ --
  although having noticeable tails for finite systems -- reduces to
  two delta functions in the thermodynamic limit.}, reflecting the
presence of many different and almost degenerate equilibrium phases,
one has that for any couple of equilibrated configurations their
mutual overlap is spatially homogeneous on a large scale.  Two
configurations with a given value of the overlap in a given
macroscopic portion of space will have the same overlap everywhere.

Starting from this observation a pair of identical systems subject to
``twisted overlap'' boundary conditions was considered in \cite{FPV}.
Two different values $q_1$ and $q_2$ of the overlap were imposed on the two
boundaries along a selected direction, chosen among the ones with non-zero
probability density in the $p(q)$.  It was then argued, that if RSB is
present, the boundaries induce a smooth overlap profile in space that
interpolates between the two values. Above the lower critical dimension
$D_c$ for spin-glass order this has a diverging free-energy cost, and one
can expect the following scaling with $L$ and $|q_1-q_2|$:
\begin{eqnarray}
  \Delta {\tilde F}(q_1,q_2) \sim L^{\theta_q} |q_1-q_2|^{b_q},
  \label{uno}
\end{eqnarray}
where the exponent $\theta_q=D-D_c$ is positive for $D>D_c$. A mean-field
calculation of $b_q$ and the interface exponent $\theta_q$ gave the
non-trivial values of $b_q=5/2$ and $D_C=5/2$. Remarkably, while measuring a
quite different property, the resulting value of the lower critical
dimension is in very good agreement with the one estimated by the
periodic-antiperiodic boundary condition method \cite{Bray-Moore,Boettcher}.

Unfortunately, despite its potentially informative content on the nature of
the spin-glass phase, the definition of \cite{FPV} was never used in
numerical studies of the spin-glass phase. This is probably due to the
difficulty of imposing values of the overlap at the boundaries.

In this paper we start from the general observation, in fact not at all
specific to spin glasses, that it is possible to induce interfaces with an 
alternative procedure. One can divide the physical system into two
contiguous halves and impose fixed values to the bulk order parameters in 
each of the two halves. This procedure, theoretically equivalent to fixing 
heterogeneous boundary conditions, is better suited for numerical
investigations, since the resulting free-energy cost can be related to the 
probability of a spontaneous fluctuation in an unconstrained system.

In the case of spin glasses we have to consider two real replicas and
impose values $q_1$ and $q_2$ of their mutual overlaps in two
contiguous half spaces.  This way of imposing heterogeneity in the
system induces an overlap profile in space equal to the one induced by
fixing the overlap on the boundaries to suitable values $q_1'$ and
$q_2'$. One then has a free-energy cost $\Delta F(q_1,q_2)$ of the
form (\ref{uno}) with the same exponents $\theta_q$ and
$b_q$.\footnote{This is strictly true for a couple of systems with
  open boundary conditions where the first procedure is naturally
  defined. The second procedure has the advantage of being defined
  also for periodic boundary conditions.}  The free-energy cost for
imposing the different overlaps is related by Boltzmann's relation to
the probability that a spontaneous fluctuation of a couple of
unconstrained systems produce the values $q_1$ and $q_2$ of the
overlaps in the two half spaces: $P_L(q_1,q_2)= e^{-\beta \Delta
  F(q_1,q_2)}$, which is a large deviation formula. This quantity has
the advantage that it is easily accessible in numerical simulations
and therefore we base our analysis on it.

In this paper we focus our attention on overlap interfaces in Dyson-like
hierarchical spin-glass models.

Hierarchical models without disorder \cite{Dyson} have played an important
r\^{o}le in the theoretical and mathematical understanding of critical
points \cite{Bleher,Jona}.  In the ferromagnetic Dyson model one can write
exact renormalization group equations that involve the iteration of a
function of a single variable (for a review see \cite{Meurice}).  
Hierarchical spin-glass models of the Dyson kind have not to our knowledge 
been considered in the literature and are in our opinion very attractive 
as they could allow for an analytical study of non-mean-field disordered 
critical points and nontrivial low temperature phases. These models provide 
a hierarchical version of spin glasses with power-law interactions
introduced in \cite{KAS}. One-dimensional models with power-law interactions 
have recently received a great deal of attention as test grounds for theoretical 
ideas about finite-range spin-glass phases \cite{Leuzzi,KY1,KY2,KY3,LPRR}. 
Here, following Dyson, we propose to use the tree topology as a further 
simplification. We choose to start the study of these models focusing on 
interface properties, for which we can get particularly simple theoretical 
predictions in the RSB framework.  We study the system through the replica
method, deriving a recursion equation relating the replica partition 
functions at the different levels of the hierarchy. This relation, which 
in principle codes for all the thermodynamic properties of the system is 
analyzed in a self-consistent way to obtain the interface free-energy. We
test our prediction in numerical simulations, finding good agreement.

The organization of the paper is the following: in section \ref{sec:general}
we discuss the definition of the interfaces and how in principle interfaces
can be evaluated and some expectations based on RSB theory.  Section
\ref{sec:hierarchical} is devoted to the definition of the models we use. In
section \ref{sec:interfaces} we discuss possible scenarios and expectations
and in section \ref{sec:derivation} we sketch the theoretical derivation of
the interface free-energy cost in hierarchical models. Section
\ref{sec:numerics} discusses the numerical simulations. Finally, section
\ref{sec:summary} concludes the paper. In the appendix we discuss the
numerical characterization of the critical point.

\section{Generalities}
\label{sec:general}

In this section we define the probability distribution of overlaps in
contiguous half spaces and discuss how it can be computed in principle in
the large deviation regime. For definiteness we consider a spin-glass system
with Ising spins $S_i=\pm 1$ on a set of indices $i\in\Lambda$ with $N$
elements, that we divide in two ``half spaces'' $\Lambda_1$ and $\Lambda_2$
with $N/2$ elements each. For two spin configurations ${\bf S}$ and ${\bf
  S'}$ one can define the partial overlaps $Q_r(S,S')$ in each of the half
spaces $\Lambda_r$, $r=1,2$ as $Q_r(S,S')=1/(N/2)\sum_{i\in \Lambda_r}S_i
S_i'$.  The joint probability distribution function (PDF) of the two
overlaps, for fixed value of the system size $L$ and quenched disorder $J$,
is given by
\begin{eqnarray}
  \hspace{-.5 cm}
  P_{J,L}(q_1,q_2)=\frac{1}{Z^2} \sum_{S,S'} 
  e^{-\beta H(S)-\beta H(S')}
  \delta(Q_1(S,S')-q_1) \delta(Q_2(S,S')-q_2),
  \label{p12}
\end{eqnarray}
where $Z$ denotes the partition function. From this relation one gets the
usual PDF of the total overlap by simple integration. One can consider the
average over the disorder of (\ref{p12}), however, here we will concentrate
on the large deviation regime, where one can expect
\begin{eqnarray}
  P_{J,L}(q_1,q_2)\sim e^{-L^{\theta_q} \Delta F_J(q_1,q_2)}
  \label{joint}
\end{eqnarray}
with a positive $\Delta F_J(q_1,q_2)$. The exponent $L^{\theta_q} \Delta
F_J(q_1,q_2)$ represents an interface free-energy cost to maintain the
constrained values. In this regime, the large deviation functional can be
expected to be self-averaging and one needs to compute the average
free-energy
\begin{eqnarray}
  E \log\left( P_{J,L}(q_1,q_2)\right) = -{L^{\theta_q} \Delta F(q_1,q_2)}. 
  \label{log}
\end{eqnarray}
The RSB implies that if one chooses $q_1=q_2$ in the domain where $p(q)$ is
non-zero, then $\Delta F(q_1,q_1)=0$. The property of homogeneity of the
overlap should then translate in a form for $\Delta F(q_1,q_2)$ of the kind
$\Delta F(q_1,q_2)=g((q_1+q_2)/2)\; |q_1-q_2|^{b_q}$.

From the complete knowledge of the joint distribution (\ref{joint}) one can
in principle extract the marginal distribution of the difference $u=q_1-q_2$
\begin{eqnarray}
  P_L(u)=\int dq_1\; dq_2\; \delta\left( u-(q_1-q_2)\right) P_L(q_1,q_2).
\end{eqnarray}
In the large deviation regime, where this analysis is supposed to be
valid, we have $|q_1-q_2| = O(1)$ for $L\to\infty$ and hence the
integral should be dominated by the value of $q=(q_1+q_2)/2$ that
maximizes the function $g$.  As a consequence $P_L(u)$, the PDF of
$u$, should behave as $e^{-g^* L^{\theta_q} |u|^{b_q}}$ in the tails.
In order to understand the behavior for ``small'' values of $u$ we can
suppose a smooth cross-over between the small and the large
fluctuation regimes. In this case the form (\ref{log}) suggests a
finite probability for $|q_1-q_2|=O(L^{-\theta_q/b_q})$, i.e., a
scaling form $P_L(q_1,q_2)=H((q_1+q_2)/2,(q_1-q_2)L^{\theta_q/b_q})$
and the marginal probability of the difference $u=q_1-q_2$,
$P_L(u)=P(uL^{\theta_q/b_q})L^{\theta_q/b_q}$. This quantity $P_L(u)$
is particularly easily accessible in numerical simulations and
contains the the information on the ratio between the interface
exponents. The comparison of this method with the study of the overlap
correlation function \cite{CGPV}, as well as the comparison of our
analytic results in section 5 with pertubative calculations \cite{DC} will be
not attempted in this paper.

\section{Hierarchical Spin Glasses}
\label{sec:hierarchical}

\begin{figure}
  \resizebox{0.75\columnwidth}{!}{%
    \includegraphics{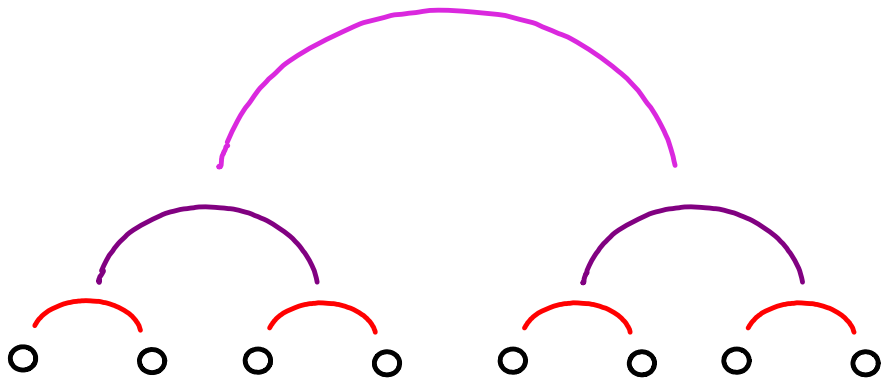}
  }
  \caption{In our models the spins are associated to the leaves of a
    binary tree with $k$ generation. In the first model all the spins interact with each other with a strength that decreases as a power of the distance. In the second case, each spin interacts directly with a small number of other spins at random, with a probability of interaction that behaves as a decreasing power of the distance.}
  \label{figuno}
\end{figure}

Here we introduce hierarchical spin-glass models on Dyson lattices.
We will consider two families of models, a first one that is better
suited for analytic studies and a second one that is more adapted to
numerical simulations.  In both families the spins are associated to the
leaves of a binary tree, see fig. (\ref{figuno}). The distance between 
two spins that, rising up in the hierarchy, meet after $l$ branches 
is naturally defined as $2^l$.  
The first family of models we would like to define, is the natural 
Spin-Glass generalization of the Ferromagnetic Dyson model (\cite{Dyson}). 
The Hamiltonian can be constructed iteratively, connecting the
two non-interacting systems of $2^k$ spins $(k \in \mathbb{N})$ to 
form a composite system of $2^{k+1}$ spins in the following way:
\begin{eqnarray}
  H_{k+1}^J[S_1,....,S_{2^{k+1}}] &=
  H_{k}^{J_1}[S_1,....,S_{2^{k}}]+H_{k}^{J_2}[S_{2^k+1},....,S_{2^{k+1}}] - \nonumber\\
  &- \frac{1}{2^{(k+1)\sigma}}\sum_{i<j}^{1,2^{k+1}}J_{ij}S_i S_j
  \label{I}
\end{eqnarray}
having defined the Hamiltonian for a system at the first level of the
hierarchy
\begin{eqnarray}
  H_1^J(S_1,S_2)&=-\frac{1}{2^{\sigma}}J S_1 S_2.
\end{eqnarray}
The couplings $J_{ij}$ are independent and identically distributed
Gaussian random variables with zero mean and unit variance. The family
is parametrized by the value of $\sigma$ that tunes the decrease of
the strength of the interaction with the distance.  The interaction
strenght decrease as a power of the distance, the model is therefore a
hierarchical counterpart of the one-dimensional spin glass with
power-law interactions \cite{KAS} which has received attention
recently \cite{Leuzzi,KY1,KY2,KY3,LPRR}. As in this case, the model is
defined for $\sigma\geq 1/2$.  Exactly at the value $\sigma=1/2$ it
reduces to the infinite range model \cite{SK}.
Notice that the sum of the squares of the interaction terms that couple the
two subsystems $ \frac{1}{2^{(k+1)2\sigma}}\sum_{i<j}^{1,2^{k+1}}J_{ij}^2 $
scales as $2^{2(k+1)(1-\sigma)}$.  This is on the order of the volume for
the mean-field value $\sigma =1/2$ and smaller than the volume for the
nontrivial regime $\sigma >1/2$.

The second family we consider is a dilute version of model (\ref{I}). 
We consider a Hamiltonian for $2^k$ spins
with a fixed number of terms $P=\alpha 2^k (\alpha > 1)$
\begin{eqnarray}
  H_{k}^J[S_1,....,S_{2^{k}}]=-\sum_{\mu=1}^P J^\mu S_{i^\mu} S_{j^\mu},
  \label{II}
\end{eqnarray}
where the $J^\mu$ are taken as $\pm 1$ variables with equal probability, and
we choose the interacting couples of sites $(i^\mu,j^\mu)$ independently
term by term and in a way that, if the distance between $i^\mu$ and $j^\mu$
on the binary tree with $2^k$ branches is $2^n$, we put the coupling with a
probability given by
\begin{eqnarray}
  P(i^\mu,j^\mu)=A_k\;
  {2^{n(1-2\sigma)}},
\end{eqnarray}
where $A_k$ is a constant chosen such that it normalizes the
probability.  For $\sigma=1/2$ the model reduces to the Viana-Bray
spin-glass model \cite{VB} on Erd\"os-R\'enyi random graph. We will
refer to the two models defined above as model I (\ref{I}) and model
II (\ref{II}), respectively.  Both models should belong to the same
universality class for the same value of $\sigma$, with respect to the
critical and the low temperature properties.  For $\sigma \in [1/2,1]$
the models have a finite temperature spin-glass transition. The
critical point has a classical (i.e., Gaussian) character for $\sigma
\in [1/2,2/3]$, while it has a non-classical character for $\sigma \in
[2/3,1]$.  As observed in \cite{LPRR,FranzParisiEPL}, diluted models
as the one in (\ref{II}) are convenient in numerical simulations since
the number of interactions for each spin does not grow with the system
size.  Spending an equal amount of computational effort we can
therefore study much bigger sizes than for model I, hoping that the
finite-size corrections are comparable in both models.

\section{Interfaces in hierarchical models}
\label{sec:interfaces}

Let us discuss some scenarios for the behavior of the PDF $P_L(q_1,q_2)$
corresponding to different possible physical situations.

The simplest physical situation is the paramagnetic state where long-range
order is absent. In this case we have a finite correlation length $\xi$ and
for large $k$ one can expect $q_1$ and $q_2$ to be sums of $\sim L/\xi$
independent terms. The resulting probability distribution of $q_1$ and $q_2$
is a product of two independent Gaussians with variance proportional to
$\xi/L$.

We can then consider a condensed phase with only two pure states. In this
case, the space average value of the overlap can take two values $\pm q^*$.
Small fluctuations will still be Gaussian as in the paramagnetic case.  Due
to the symmetry of the model, where groups of spins at a given level of the
hierarchy are on the same foot, one can expect that large fluctuations where
$q_1\approx q^*$ and $q_2\approx -q^*$ (or vice versa) will imply a
free-energy cost $\Delta F=A 2^{(k+1)2(1-\sigma)}$ and have a probability
$e^{-\beta A 2^{(k+1)2(1-\sigma)}}$.

The most interesting possibility is a spin-glass phase with RSB. In
this case one has a zero mode in the free-energy associated to the
existence of a couple of states with overlaps taking values in a
finite interval \cite{MPV}. If one chooses $q_1=q_2=q$ one finds a
broad distribution $p_{k+1}(q)$, which for large $k$ is close to the
limiting distribution $p(q)$. Consider first a system in absence of
interactions at the $k+1$-th level. The two subsystems are independent
and $P^{(0)}_{k+1}(q_1,q_2)=p_k(q_1)p_k(q_2)$.  If the interaction is
switched on, we can expect a free-energy cost $\Delta
F(q_1,q_2)=g(q_1+q_2) 2^{2(1-\sigma)k}|q_1-q_2|^{b_q}$.  The overlap
interface exponent is just given by $\theta_q=2(1-\sigma)$ due to the
fact that the total interaction strength {\it squared} between the two
parts scales as $2^{2(1-\sigma)k}$.  The value of the exponent $b_q$
as well as the function $g$ can be computed supposing RSB at the level
$k$.  The detailed calculation is quite involved and is presented in
the next section; here we just give the net result, valid in the
regime $q_1,q_2\in [-q_{EA},q_{EA}]$.  Neglecting prefactors we could
not compute, it reads:
\begin{eqnarray}
  P_{k+1}(q_1,q_2) \sim e^{-p((q_1+q_2)/2) 2^{2(1-\sigma)(k+1)}|q_1-q_2|^3}.  
  \label{main}
\end{eqnarray}
Notice the appearance of the $k\to\infty$ function $p(q)$ in the exponent of
(\ref{main}).  From this formula one can extract the conditional probability
of the difference $u=q_1-q_2$ for a fixed value of the semi-sum
$q=(q_1+q_2)/2$:
\begin{eqnarray}
  P_{k+1}(u|q) \sim e^{-p(q) 2^{2(1-\sigma)(k+1)}|u|^3}.  
  \label{diff}
\end{eqnarray}
Equation (\ref{diff}) summarizes our prediction for the hierarchical model.
For a fixed value of the sum, the difference of the overlap is distributed
according to the exponential of the cube, which is different from a naive
Gaussian guess. The coefficient of the exponential is equal to the function
$p(q)$ which can be evaluated in independent measurements.  In numerical
simulations it might be complicated to collect sufficient statistics to
condition $u$ to the value of $q$.  One can then turn to the unconditional
distribution. In the large deviation regime, where $u \sim O(1)$, this should be
dominated for large $k$ and finite $u$ by the value of $q$ which maximizes
(\ref{diff}). For functions $p(q)$ as the one commonly met in spin-glass
systems this is the value $q=q_{EA}$ which is the largest possible value of
the overlap in the thermodynamic limit.

The form (\ref{diff}) suggests that the order of magnitude of the typical
fluctuations is $|u|\sim 2^{2/3(1-\sigma)(k+1)}$. In this regime, the
unconditional distribution of $u$ involves the convolution of (\ref{diff})
with a presently unknown prefactor and cannot be computed. All the form
(\ref{diff}) tells us for the unconditioned distribution in this regime is
that the finite volume distribution of $u$ admits the scaling form
\begin{eqnarray}
 \hspace{-1.5 cm}
 P_{k+1}(u) =\int dq P_{k+1}(u|q)P_{k+1}(q)=W(2^{2/3(1-\sigma)(k+1)}|u|)2^{2/3(1-\sigma)(k+1)}.
  \label{uncond}
\end{eqnarray}
In order to compute the scaling function $W$ the knowledge of the prefactor
in (\ref{diff}) would be necessary.  We remark that the exponent in this
scaling is a strong consequence of RSB theory, a naive guess would have
suggested a Gaussian distribution with scaling variable
$2^{2(1-\sigma)(k+1)}u^2$.

\section{Replica derivation of formula (\ref{diff}) }
\label{sec:derivation}

We analyze model I through the replica method. In order to compute the
free-energy, it is natural to consider a recursion that relates the average
partition function of $n$ replicas $S_i^a$, $a=1,...,n$, with fixed mutual
overlaps $Q_{ab}=\frac{1}{2^k}\sum_{i=1}^{2^k} S_i^a S_i^b$.  Defining
\begin{equation}
  \fl Z_{k}[{\bf Q}] = 
  E_J\left[\sum_{{\bf S}}\exp\left(-\sum_{a=1}^n
      H_k^{J}(S_1^a,...,S_{2^k}^a)\right)\prod_{a<b}^{1,n}\delta
    \left(Q_{ab}-\frac{1}{2^k}\sum_{i=1}^{2^k} S_i^a S_i^b \right)\right]\!,
\end{equation}
where $E_J$ denotes the average over the disorder, we can write:
\begin{equation}
  \fl Z_{k}[{\bf Q}] = 
  \exp\left( \frac{\beta^2}{4}2^{2(1-\sigma)k} \Tr {\bf Q}^2 \right) \int
  {\cal D} {\bf Q}_1{\cal D} {\bf Q}_2 Z_{k-1} [{\bf Q}_1]Z_{k-1}[{\bf
    Q}_2]\delta \left( {\bf Q}-\frac{{\bf Q}_1+{\bf Q}_2}{2} \right)\!.
  \label{zeta}
\end{equation}
For integer $n$ this is an exact relation based on the independence of the
Hamiltonians of the sub-systems at level $k$.  In principle, the
thermodynamical properties of the system are encoded in this recursion and
in its analytic continuations for $n \to 0$. For example it can be used to
set up an epsilon expansion for the calculation of the critical indexes for
$\sigma \in [2/3,1]$ \cite{workinprog}.  In this paper we just use
(\ref{zeta}) to study the probability distribution (\ref{diff}). This can be
done with the technique of constrained replicas, introduced and discussed at
length in \cite{FPV}. In order to consider constrained free-energies for two
replicas with fixed overlaps one should fix some of the elements of the
matrix $Q_{ab}^r$ $r=1,2$ to the values of the constraints. Writing $n=2n'$
the constraint reads $Q_{a,a+n'}^r =Q_{a+n',a}^r=q_r$ for $r=1,2$ and
$a=1,...,n'$.  Introducing the replica free-energy at level $k$ for fixed
$q_1$ and $q_2$, $Z_k[{\bf Q}|q_1,q_2]=e^{-\beta 2^k F_k[{\bf Q}|q_1,q_2]}$
we see that one needs in principle to compute:
\begin{eqnarray}
 \fl \int {\cal D}{\bf Q}\; e^{-\beta 2^{k} F_{k}[{\bf Q}|q_1,q_2]} =& \int {\cal
    D} {\bf Q}_1{\cal D} {\bf Q}_2 e^{\left(
      \frac{\beta^2}{4}2^{2(1-\sigma)k} \Tr ({\bf Q}_1+{\bf Q}_2)^2 \right)}
  e^{-\beta 2^k (F_{k}[{\bf Q}_1]+F_k[{\bf Q}_2])} \times \nonumber \\
  &\times \prod_{a=1}^{n'}\delta(Q^1_{a,a+n'}-q_1)\delta(Q^2_{a,a+n'}-q_2).
  \label{schif}
\end{eqnarray}
This form suggests that for large $k$, supposing the knowledge about
$F_{k}[{\bf Q}]$, the integral over ${\bf Q}_1$ and ${\bf Q}_2$ can be
performed by the saddle-point approximation.  Notice that the interaction
term is sub-extensive and scales as $2^{2(1-\sigma)(k+1)}$, while the
partial free-energies scale as the volume $2^k$. The interaction term does
therefore not contribute to the saddle point, and the maximization with
respect to the matrices $Q^r$ can be performed separately in each of the two
sub-systems.

The value of the interaction term at the saddle point determines the
free-energy difference $\Delta F(q_1,q_2)$.

We now study the consequences of the hypothesis that there is RSB in the
system. Specifically, we suppose that in absence of any constraints RSB is
described by a continuous Parisi function $q_F(x)$ taking values between the
two extremes $-q_{EA}$ and $q_{EA}$. In the constrained problem,
correspondingly, each of the matrices $Q^r$ is parametrized by two functions
$q^r(x)$ and $p^r(x)$ with $x \in [0,1]$. As implied by the analysis in
\cite{FPV} the free-energy in each of the two sub-systems is then
independent of $q^r$, and the function $q^r(x)$ and $p^r(x)$ can be directly
related to the function $q_F(x)$ of the unconstrained system by the
relations
\begin{eqnarray}
  \hspace{-1.7 cm}
  q(x)=\left\{ 
    \begin{array}{cc}
      q_F(2 x) &\;\;\; x\leq {x_1}/2\\
      q_r& \;\;\;  x_r/{2}< x\leq x_r\\
      q_F(x) &\;\;\; x> x_1
    \end{array}
    \;\;
  \right.
  p_x(u)=\left\{ 
    \begin{array}{cc}
      q_F(2 x) &\;\;\; x\leq {x_r}/2\\
      q_r& \;\;\;  x> x_r/{2},
    \end{array}
  \right.
  \label{p2}
\end{eqnarray}
where $x_r$ is the value of $x$ such that $q_F(x_r)=q_r$.  We then see that
the free-energy difference from the unconstrained case is entirely due to
the interaction term, which can be evaluated using the saddle-point value of
the matrices $Q_1$ and $Q_2$:
\begin{equation}
\fl  \Tr \left( \frac{{\bf Q}_1+{\bf Q}_2}{2}\right)^2
  = n\left[\left(\frac{q_1+q_2}{2}\right)^2- \int dx \left(\frac{q_1(x)+q_2(x)}{2}\right)^2+
    \left(\frac{p_1(x)+p_2(x)}{2}\right)^2 \right]\!\!.
\end{equation}
Substitution of (\ref{p2}) leads to
\begin{equation}
 \fl \Tr \left( \frac{{\bf Q}_1+{\bf Q}_2}{2}\right)^2 
  = n \left( -2\int_0^1 dx q_F(x)^2+\int_{x_1}^{x_2}(q_2-q_F(x))(q_F(x)-q_1)\right)\!,
\end{equation}
where without loss of generality we have supposed $q_1<q_2$.  The first term
is just the contribution that can be expected if $q_1=q_2$. Together with
the two subsystems' free-energy it just gives the free-energy of the system
at the level $k+1$. The second term is associated to the free-energy excess
needed to impose $q_1\ne q_2$.  For small $q_2-q_1$ we can expand this last term
and find
\begin{eqnarray}
  n x'\left(\frac{q_1+q_2}{2}\right) (q_2-q_1)^3\!,
\end{eqnarray}
where $x(q)$ is the inverse function of $q_F(x)$. Inserted in (\ref{schif})
and identifying $x'(q)$ with $p(q)$ as discussed in length in \cite{MPV}
leads to (\ref{diff})
\begin{eqnarray}
  P_{k+1}(q_1,q_2) \sim e^{-p((q_1+q_2)/2) 2^{2(1-\sigma)(k+1)}|q_1-q_2|^3}\!\!.  
  \label{main1}
\end{eqnarray}

\section{Numerical Simulations}
\label{sec:numerics}

In this section we discuss the results of numerical simulations used to test
the behavior of the probability distribution of the overlap difference in
the two subsystems. In order to deal with PDF's of a single variable we
concentrated to the unconditional probability for which we have the
theoretical prediction (\ref{uncond}).  We did not try to test the more
detailed prediction (\ref{diff}) which would need the numerical
determination joint PDF's of two variables.

We have simulated model II for $\alpha=3/2$, using parallel tempering to
thermalize the system at low temperatures.

We have tested our theoretical predictions concentrating on two values of
$\sigma$: $\sigma=0.576$ which lies in the classical region $1/2<\sigma<2/3$,
where the spin-glass transition is well described by mean-field theory, and
$\sigma=0.707$ which lies in the non-classical region $2/3<\sigma<1$, where
the exponents are nontrivial.

We have characterized the critical point of the model following the
procedure that we describe in the appendix. We estimate the critical
temperatures to be equal to $T_c=1.24(1)$ for $\sigma=0.576$ and $T_c=1.14(1)$ for
$\sigma=0.707$.

To test the predictions described in the previous sections we measured the
distribution of the variables $q$ and $u$ as a function of temperature and
system size. We considered systems sizes of $2^7, 2^8, 2^9$ and $2^{10}$
spins. Averages were performed over 4000 samples for the smaller systems and 
700 samples for the largest system. The configurations are thermalized
during the first $2^{20}$ Monte Carlo (MC) sweeps of the runs and then data 
is collected for the subsequent $2^{20}$ MC steps.

\begin{figure}
  \resizebox{0.75\columnwidth}{!}{%
    \includegraphics{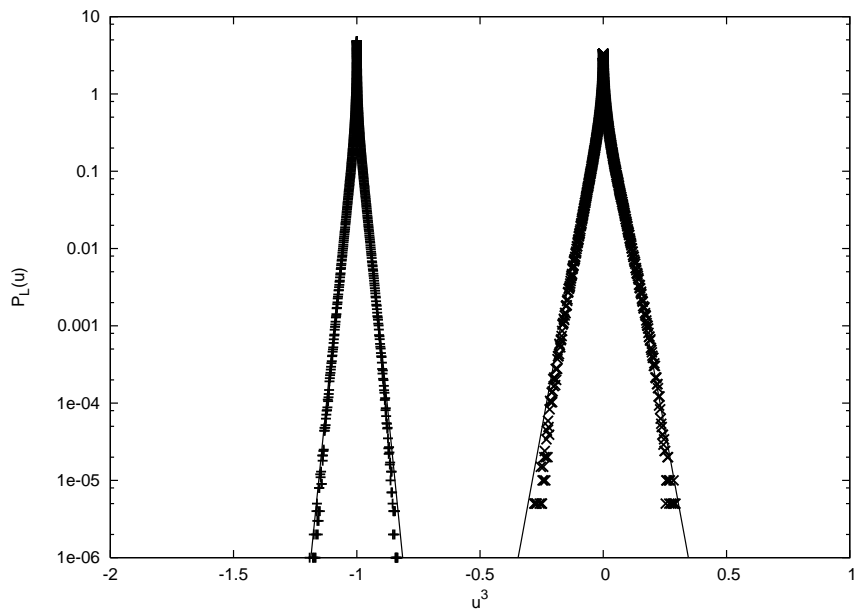}
  }
  \caption{Probability distribution function $P_k(u)$ for $K=10$,
    $\sigma=0.576$ (left) in the classical region and $\sigma=0.707$
    (right) in the non-classical region, plotted in a log-linear scale
    as a function of $u^3$. Both systems are in the low temperature
    phase, the temperatures being respectively $T=0.745$ and
    $T=0.740$.  The data for $\sigma=0.576$ have been shifted of one
    unity to the right. For comparison we plot curves of the kind $a\;
    e^{b u^3}$ with $a$ and $b$ adjusted to fit the tails of the
    curves. }
  \label{fig1}
\end{figure}

The first prediction we test is the validity of the cube-exponential form in
the tails for large $u$. This is well observed in all our simulations. A
typical example of our findings is depicted in figure (\ref{fig1}) where we
plot the function $P_k(u)$ for $k=10$ for the two different values of
$\sigma$ in the classical and non-classical regime, respectively. The data
are plotted together with functions of form $f(x)=a\; e^{-b|x|^3}$ which
should be considered as a guide to the eye rather than the best fit. The
parameters $a$ and $b$ were fixed by eye to be equal to $a=1$, $b=40$ for
$\sigma=0.707$ and $a=0.8$, $b=73$ for $\sigma=0.576$. A best fit procedure
results to be sensitive to the chosen fitting interval and to the tails of
the distribution that represent probabilities too small to be correctly
estimated with our statistics. Despite these caveats we believe that our
data give an indication in favor of the cubic behavior for both values of
$\sigma$.

\begin{figure}
  \resizebox{0.75\columnwidth}{!}{%
    \includegraphics{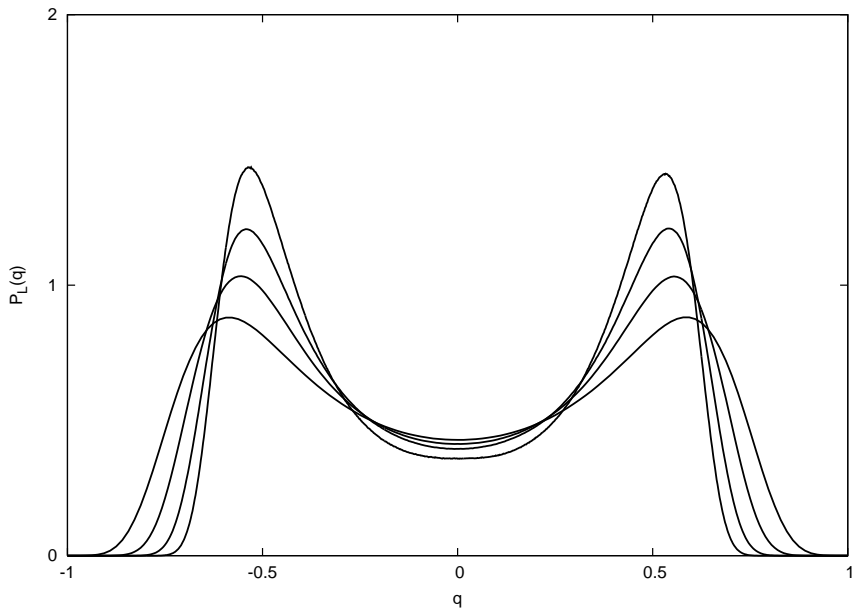}
  }
  \caption{Probability distribution function $p(q)$ for $\sigma=0.576$
    (classical region) and $T=0.745=0.615T_c$ ($T_c=1.24(1)$) for $L=2^k$ with
    $k=7,8,9,10$. The peaks become sharper and sharper for larger $k$.  }
  \label{fig2} 
\end{figure}

\begin{figure}
  \resizebox{0.75\columnwidth}{!}{%
    \includegraphics{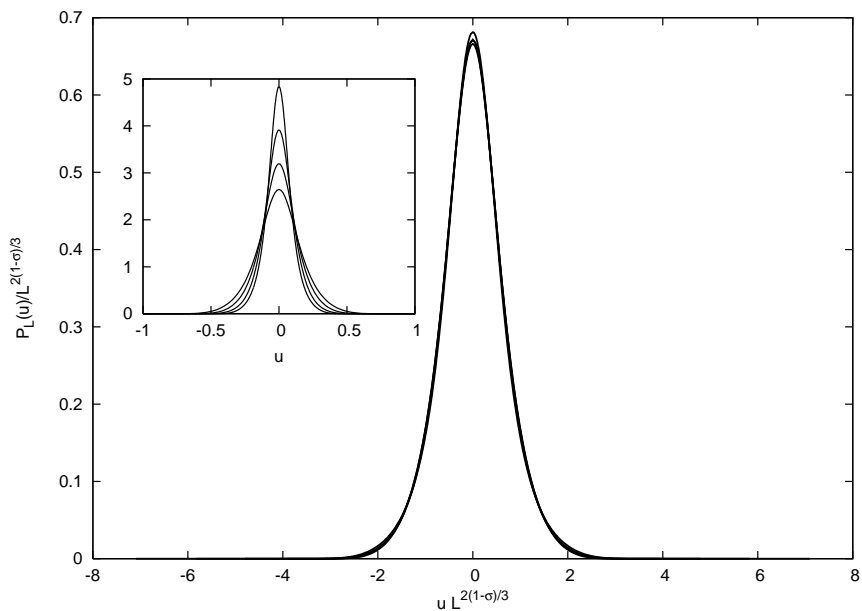}
  }
  \caption{Scaling plot of probability distribution of the overlap
    difference for $\sigma=0.576$ (classical region) and $T=0.745=0.601T_c$
    for $L=2^k$ with $k=7,8,9,10$. Inset: the unscaled probability.}
  \label{fig3} 
\end{figure}

\begin{figure}
  \resizebox{0.75\columnwidth}{!}{%
    \includegraphics{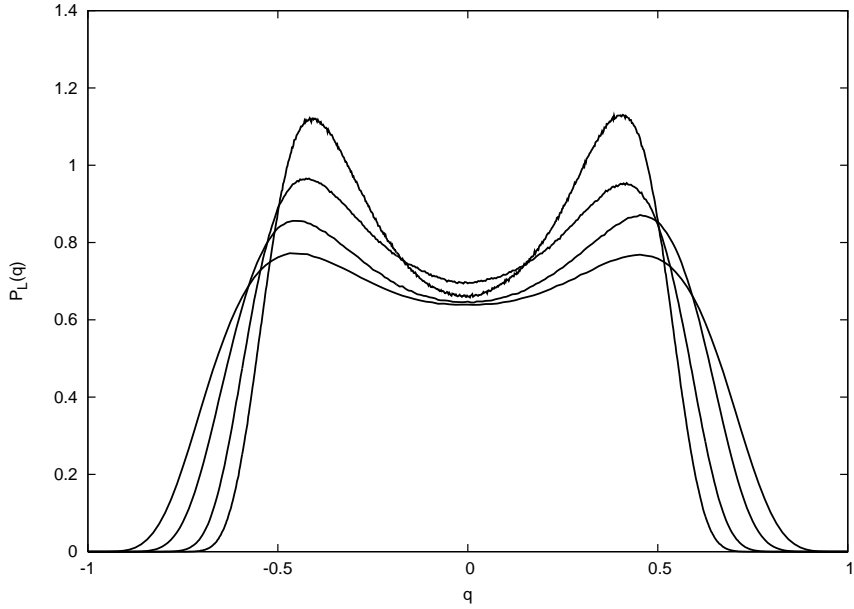}
  }
  \caption{Same as figure \ref{fig2} for $\sigma=0.707$ in the non-classical
    region. Here $T_c=1.14(1)$ and we present data for $T=0.740=0.65 T_c$. }
  \label{fig4}
\end{figure}

\begin{figure}
  \resizebox{0.75\columnwidth}{!}{%
    \includegraphics{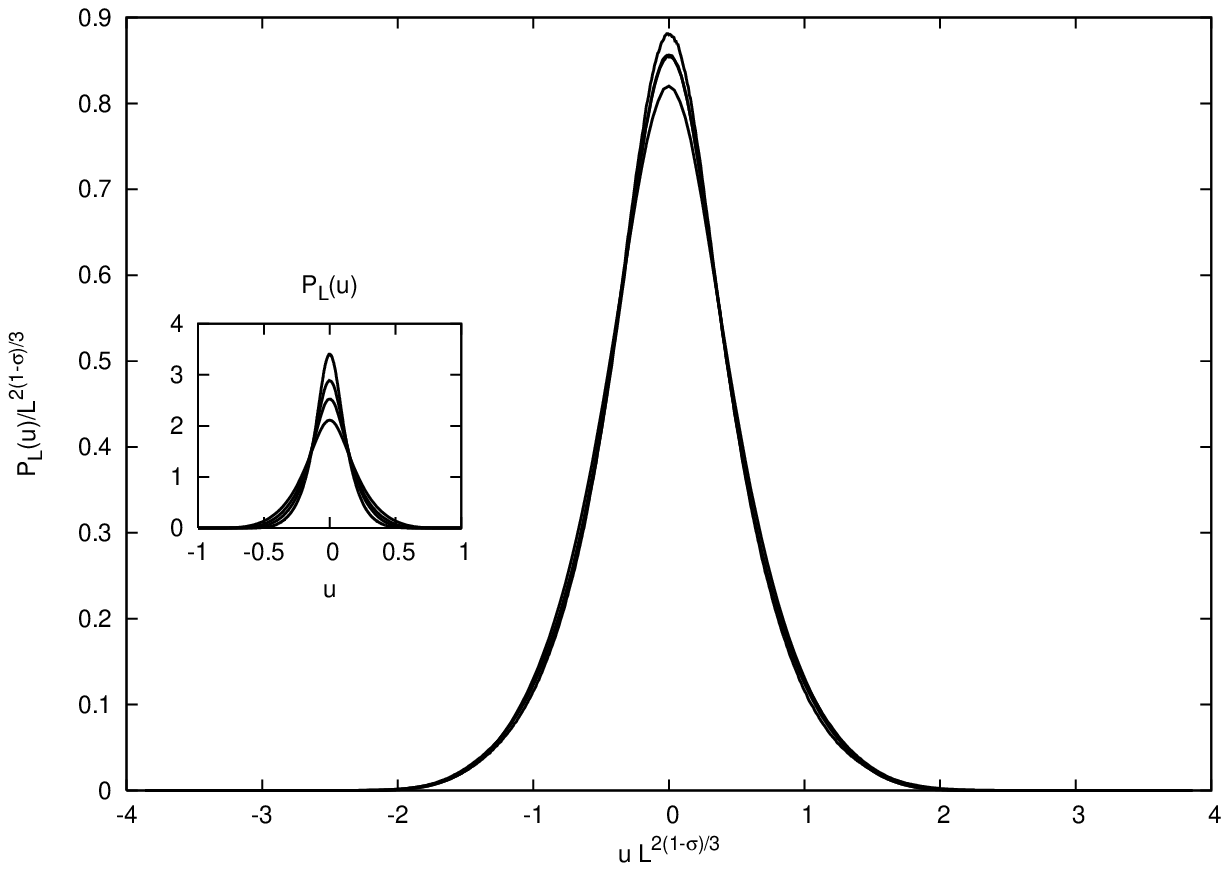}
  }
  \caption{Same as figure \ref{fig3} for $\sigma=0.707$ in the non-classical
    region. The temperature is the same as in figure \ref{fig4}, $T=0.740$. }
  \label{fig5} 
\end{figure}

We then investigated the scaling of the PDF of $u$ with the system
size, contrasting it with the behavior of the PDF of $q$. In figures
\ref{fig2} and \ref{fig4} we display the function $P_k(q)$ for the two
values of $\sigma$ at low temperatures.  The function $P_k(q)$ has the
characteristic appearance of the one of systems developing RSB for
large volumes, with two symmetric peaks and a non-zero part for
$q\approx 0$. Figures \ref{fig3} and \ref{fig5} in contrast show that
the distribution of $u$, $P_k(u)$, is unimodal around zero.  Its width
is as expected a decreasing function of $k$. In the insets the
unscaled data are presented, while the in the main panel the result of
scaling the data using the variable $u 2^{\frac 2 3 (1-\sigma)}$ is
shown.  We judge the scaling (\ref{uncond}) to be very satisfactory,
though not perfect for the values of $k$ we explored. Indeed, while
for $\sigma=0.576$ we could not find a the value of the ratio
$\theta_q/b_q$ producing a better data collapse than the theoretical
value ${\frac 2 3 (1-\sigma)}=0.28$, for $\sigma=0.707$ the value
$\theta_q/b_q=0.22$ produces a better data collapse than the
theoretical value ${\frac 2 3 (1-\sigma)}=0.19$. We believe that this
discrepancy is due to finite size effects, but we can not exclude at
present that the theory should be amended in the non-classical region.

 Our data show that
the best fitting exponent is largely temperature independent in the low
temperature region. A direct scaling of the data in the high temperature
region for the values of $k=7,8,9,10$ that we dispose, produces an effective
exponent that crosses over slowly from the low temperature value towards the
paramagnetic value $\theta_q/b_q=0.5$. We believe that we are seeing a
preasymptotic behavior due to the influence of the critical fixed point.
This influence could be particularly marked due to the power-law
interactions where the critical fixed point continues to attract the system
on relatively large scales.  Larger and larger values of $k$ are necessary
to observe the paramagnetic behavior closer and closer to the critical
temperature.



\section{Summary and Conclusion}
\label{sec:summary}

In this paper we have proposed a new method to determine numerically the
overlap interface exponents first defined in \cite{FPV}. These can be
obtained by looking at the finite-size scaling of the PDF of the differences
of the overlaps between two replicas in two subsystems. We have applied the
definition to hierarchical models, where we could give a theoretical
prediction for the scaling of the overlap differences with size.  Though the
interface exponent is naturally dictated by the model, the dependence in $u$
is found to be nontrivial in presence of RSB.  We tested this dependence in
numerical simulations finding very satisfactory agreement both in the
classical and in the non-classical region. These results confirm the
interest of hierarchical spin-glass models, that combine analytical
tractability, nontrivial critical points and RSB low temperature phases.

{\bf Acknowledgments} It is a pleasure to thank O.~C.\ Martin, M.\
M\'{e}zard, F.\ Ricci-Tersenghi, P. Contucci, C. Giardin\`a, C. Giberti
and C. Vernia for interesting discussions.

\section*{Appendix: Characterization of the critical point.}
In this appendix we discuss the characterization of the critical point of
the model for the values of the interaction parameter $\sigma$ that we have
considered in the text.

The model has been simulated with the Parallel Tempering algorithm, with 10
values of the temperature, using $2^{20}$ thermalization steps before
collecting data in the $2^{20}$ steps.  We considered systems sizes of $2^7,
2^8, 2^9$ and $2^{10}$ spins. Averages were performed over 4000 samples for
the smaller systems and 700 samples for the largest system.  Simulating two
replicas in parallel, we have studied the second and the fourth moment of
the distribution of the mutual overlap $q_2=E\langle Q^2 \rangle$ and
$q_4=E\langle Q^4 \rangle$.  We have identified the critical temperature and
the exponents $\eta$ and $\nu$ using finite size scaling through the
behavior of $q_2$, $q_4$ and the corresponding Binder parameter \cite{Binder}
$B=\frac{1}{2}(3-\frac{q_4}{q_2^2})$. In the non-classical region, $\sigma> 2/3$
where finite-size scaling should hold the various parameters exhibit the
following dependence on the temperature and system size $L=2^k$:
\begin{eqnarray}
  \label{noncl}
  &&\chi=L q_2 = L^{2-\eta} g_2(L^{1/\nu} (T-T_c))\\
  \nonumber
  &&L^2 q_4 = L^{4-2\eta} g_4(L^{1/\nu}(T-T_c))\\
  \nonumber
  &&B=b(L^{1/\nu}(T-T_c)).
\end{eqnarray}
The exponent $\eta$ should not renormalize in long range models, and 
analogously to the Euclidian 1D model take
the value $\eta=3-2\sigma$ both in the non-classical and in the classical
regions\cite{KAS}.

In the classical region, $1/2<\sigma<2/3$, the scaling implied by
(\ref{noncl}) does not hold \cite{Brezin,BZJ}.  It is possible to show,
using the fact that the critical theory is described by a cubic action
analogous to the one for short-range spin glasses \cite{HLC}, that
the various quantities scale according to the following:
\begin{eqnarray}
  \label{sc}
  &&L q_2 = |T-T_c|^{-\gamma}  {\tilde g}_2(L^{1/3}(T-T_c))\\
  \nonumber
  &&L^2 q_4 = |T-T_c|^{-2 \gamma} {\tilde g}_4(L^{1/3}(T-T_c))\\
  \nonumber
  &&B={\tilde b}(L^{1/3}(T-T_c)).
\end{eqnarray}
The exponent $1/3$ in the scaling functions can be derived from dimensional
analysis from the cubic action. The exponent $\gamma$ takes the value
$\gamma_{cl}=1$ independently of $\sigma$ as can be checked from the
scaling relation $\gamma=(2-\eta)\nu$ with $\nu=\frac{1}{2\sigma-1}$. 

\begin{figure}
  \resizebox{0.75\columnwidth}{!}{
    \includegraphics{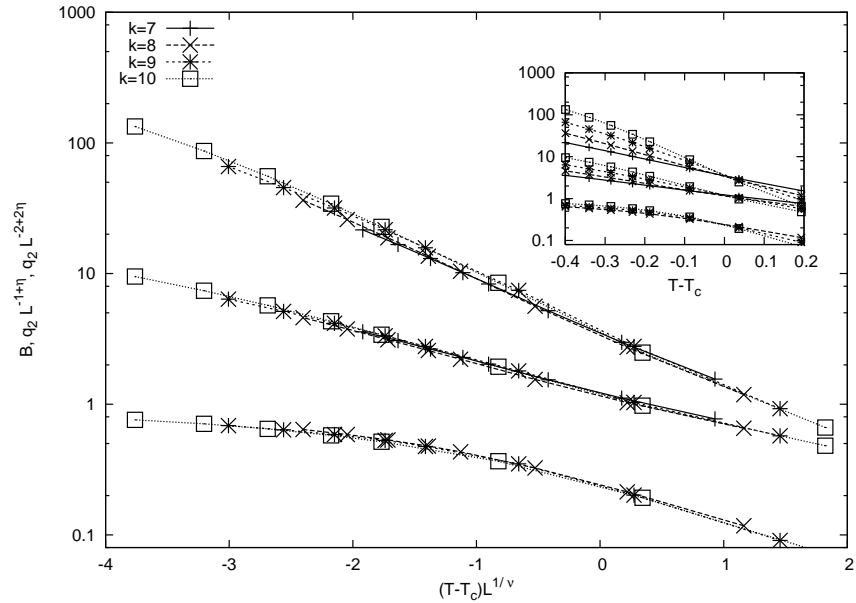}
  }
  \caption{Behavior of $q_2$, $q_4$ and $B$ as a function of temperature and
    system size for $\sigma=0.707$. In the inset we plot $q_2 /L^{1-\eta}$,
    $q_4 /L^{2-2\eta}$ and $B$ against $(T-T_c)$, with the values $T_c=1.14$
    and $\eta=3-2\sigma$.  Main panel, scaling plot of the data $q_2
    /L^{1-\eta}$, $q_4 /L^{2-2\eta}$ and $B$ against $(T-T_c)L^{1/\nu}$
    assuming the value of $\nu=3.09$. }
  \label{fig6}
\end{figure}

Let us now turn to the data considering the non-classical region
first. To analyze the data we observe the following procedure: we
first estimate the critical temperature from the crossing point of the
Binder parameter and the rescaled values, especially, $q_2
L^{-1+\eta}$ and $q_4 L^{-2+2\eta}$ since these provide for a cleaner
crossing than the Binder parameter.  We then fix the value of $\nu$ in
order to collapse the curves. The result for $\sigma=0.707$ is shown
in figure \ref{fig6}, we present the data for $q_2$, $q_4$ and $B$
scaled as in (\ref{sc}) using a value of $T_c=1.14$.

If we try to use the scaling (\ref{noncl}) in the classical regime we
get inconsistent results: although we obtain an approximate crossing
of the curves for the three quantities, the temperatures at which the
curves cross do clearly not coincide.  The crossing of the Binder
parameter indicates $T_c=1.24$ and as shown in figure \ref{fig6} for
$\sigma=0.576$, we obtain consistent scaling assuming the form
(\ref{sc}) and not (\ref{noncl}).

\begin{figure}
  \resizebox{0.75\columnwidth}{!}{%
    \includegraphics{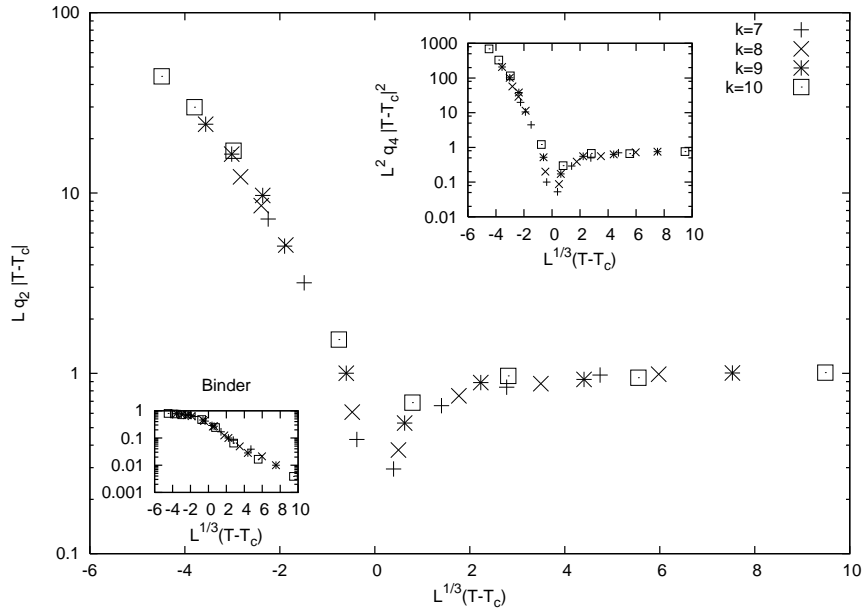}
  }
  \caption{Behavior of $q_2$, $q_4$ and $B$ as a function of
    temperature and system size for $\sigma=0.576$. In the lower inset
    we show the Binder, for which the data collapse is obtained for
    $T_c=1.14$. The main panel shows $L q_2 |T-T_C|$. Notice the high
    temperature behavior corresponding to the Curie law.  The curves
    for larges systems sizes show a tendency of converging toward the
    value 1 close the critical temperature. In the upper inset we show
    the scaling for $L^2 q_4 |T-T_C|^2$ which shows the same
    qualitative behavior. }
  \label{fig7}
\end{figure}

\section{Bibliography}
\label{sec:bibliography}

\end{document}